\pdfoutput=1
\documentclass[
superscriptaddress,
preprint,
 amsmath,amssymb,
]{revtex4-1}

\usepackage{amsmath}
\usepackage{amssymb}
\usepackage{amsfonts}
\usepackage[]{graphicx}% Include figure files
\usepackage{dcolumn}% Align table columns on decimal point
\usepackage{bm}% bold math
\usepackage{braket}
\usepackage{newtx}
\usepackage{hyperref}
\hypersetup{colorlinks=true, citecolor=green, urlcolor=blue, linkcolor=blue}
\usepackage{color}
\usepackage{ulem}

\begin{document}

%\preprint{APS/123-QED}

\title{Possible Pairing Symmetry of BaPtAs$_{1-x}$Sb$_{x}$ with an Ordered Honeycomb Network}% Force line breaks with \\

\author{Tsuyoshi Imazu}
\email[]{imazu.tsuyoshi@jaea.go.jp}
\affiliation{Department of Mathematics and Physics, Hirosaki University, 3 Bunkyo-cho, Hirosaki 036-8561, Japan}
\affiliation{Advanced Science Research Center, Japan Atomic Energy Agency, Tokai, Ibaraki 319-1195, Japan}

\author{Naoya Furutani}
\affiliation{Department of Physics, Okayama University of Science, 1-1 Ridai-cho, Kita-ku, Okayama 700-0005, Japan}

\author{Tadashi Adachi}
\affiliation{Department of Engineering and Applied Sciences,
Sophia University, 7-1 Kioi-cho, Chiyoda-ku, Tokyo 102-8554, Japan}

\author{Kazutaka Kudo}
\affiliation{Department of Physics, Graduate School of Science,
Osaka University, 1-1 Machikaneyama, Toyonaka 560-0043, Japan}
\affiliation{Institute for Open and Transdisciplinary Research Initiatives, Osaka University, 1-1 Yamadaoka, Suita 565-0871, Japan}

\author{Yoshiki Imai}
\affiliation{Department of Physics, Okayama University of Science, 1-1 Ridai-cho, Kita-ku, Okayama 700-0005, Japan}

\author{Jun Goryo}
\email[]{jungoryo@hirosaki-u.ac.jp}
\affiliation{Department of Mathematics and Physics, Hirosaki University, 3 Bunkyo-cho, Hirosaki 036-8561, Japan}

\date{\today}% It is always \today, today,

\begin{abstract}
We investigate the possible pairing symmetry of superconducting $\rm{BaPtAs}_{1-\it{x}}\rm{Sb}_{\it{x}}$ solid solution with an ordered-honeycomb network of Pt and pnictogens. 
A spontaneous internal magnetic field below the superconducting transition temperature is observed in BaPtSb ($x = 1$) via the muon-spin relaxation measurement. 
We then pursue a scenario where the pairing symmetry is changed from a time-reversal symmetry-breaking (TRSB) state to another one by changing the Sb-concentration utilizing the effective tight-binding model obtained from the first principles calculations for $x = 0$ and $x = 1$, at which we see a significant difference in the shape of the dominant Fermi surfaces.
We find that the chiral $d$-wave state with TRSB is most stable at $x = 1$, whereas the nodal $f$-wave or the conventional $s$-wave states without TRSB are competitive at $x = 0$.
\end{abstract}

\maketitle

%\tableofcontents

\section{\label{sec:level1}Introduction}

The crystal symmetry plays a crucial role in classifying the pairing states \cite{Sigrist-rmp-1991}. 
Each pairing state is categorized into an irreducible representation of the point group of the crystal lattice. 
The most stable state with the highest condensation energy (or transition temperature $T_{\rm{c}}$) is determined by the electronic structure and pairing interaction. 
In hexagonal crystal,  a two-dimensional irreducible representation comprising $d_{x^2-y^2}$- and $d_{xy}$-wave can form a chiral $d_{x^2-y^2} \pm i d_{xy}$-wave (chiral $d$-wave) state. 
The state is discussed theoretically, for instance, the superconductivity in heavily doped graphene \cite{Black-Schaffer-jpcm-2014}, but has remained elusive. 
The chiral $d$-wave state is a topological state with broken time-reversal symmetry, and the Chern number characterizes its topological nature \cite{Schnyder-prb-2008}. 
The state gives rise to the spontaneous chiral current at the sample surface or around defects.  
Such a current causes the spontaneous magnetic field below $T_{\rm{c}}$ that could be detected by muon-spin relaxation ($\mu$SR) measurement  \cite{Ghosh-jpcm-2021} .

 \begin{figure}[tb]
\includegraphics[width=\linewidth]{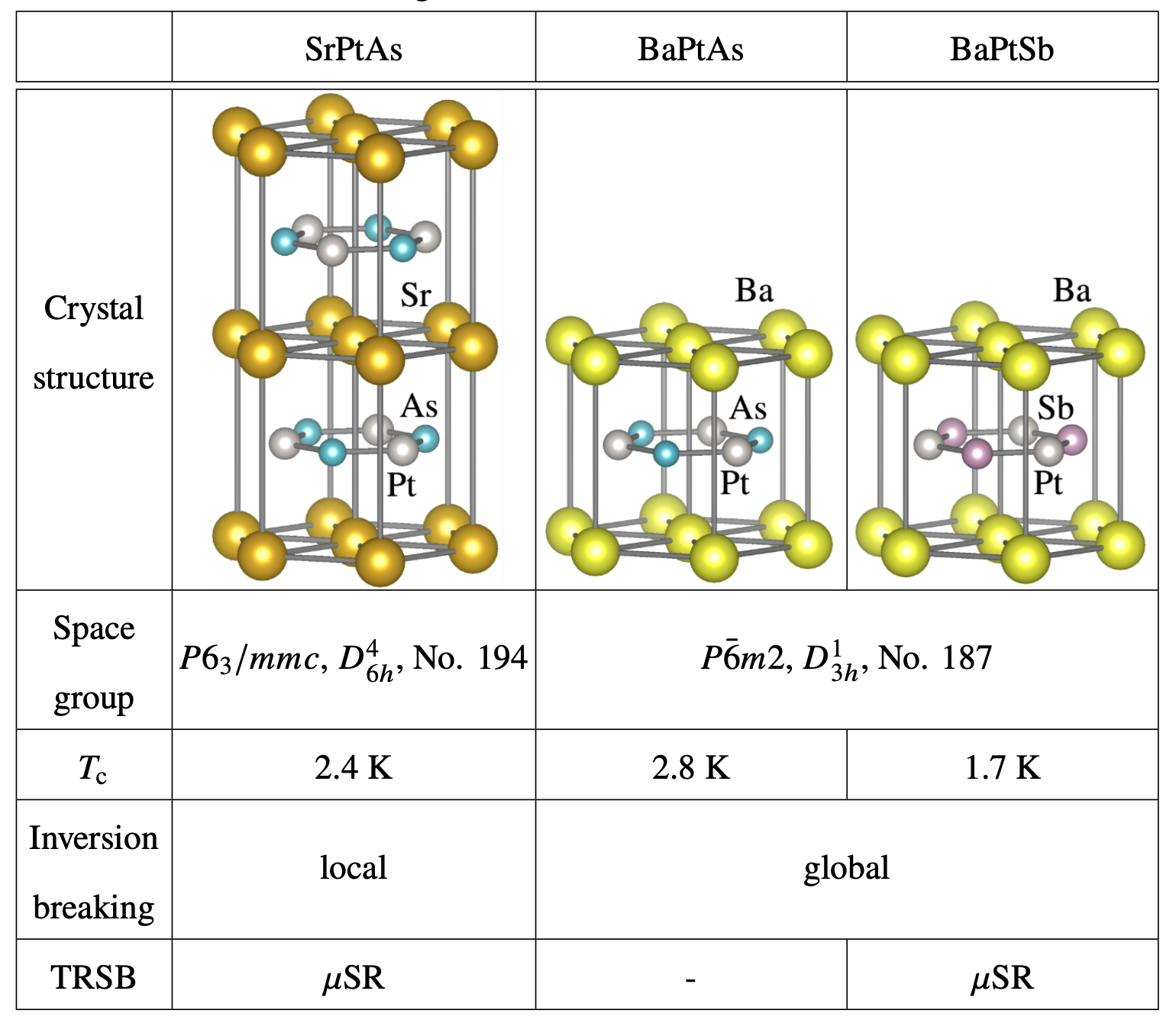}% Here is how to import EPS art
\caption{\label{fig:HoneySC} (Color online) Crystal structures and properties of ordered honeycomb network superconductors  \cite{Wenski-1986,Kudo-jpsj-2018-BaPtAs,Nishikubo-jpsj-2011,Kudo-jpsj-2018-BaPtSb,Biswas-prb-2013,Adachi-prb-2025}. The crystal structures are visualized using VESTA \cite{Momma-jac-2011}. }
\end{figure}
 
Recently, a series of pnictide superconductors with an ordered-honeycomb network has been discovered \cite{Nishikubo-jpsj-2011}.  
The spontaneous magnetization has been observed by $\mu$SR in the first discovered $\rm{SrPtAs}$ with crystal point group $D_{6h}$ (Fig.~\ref{fig:HoneySC}) \cite{Biswas-prb-2013}. 
Although some estimations support the chiral $d$-wave pairing \cite{Goryo-prb-2012,Fischer-prb-2014,Ueki-prb-2019}, the situation is controversial since several reports suggest the conventional $s$-wave or line-nodal $f$-wave states \cite{Matano-prb-2014, Landaeta-prb-2016,Bruckner-prb-2014,Wang-prb-2014}. 
BaPtAs ($T_{\rm{c}} = 2.8~\rm{K}$) and BaPtSb ($T_{\rm{c}} = 1.7~\rm{K}$) possess $D_{3h}$ symmetry (Fig.~\ref{fig:HoneySC}), and non-monotonic variation of $T_{\rm{c}}$ concerning a change of $x$ has been observed in  $\rm{BaPtAs}_{1-\it{x}}\rm{Sb}_{\it{x}}$  solid solution \cite{Ogawa-jpsj-2022}.
Intriguingly, $\mu$SR detects spontaneous internal magnetic field at $x = 1$, whereas the signal is almost (completely) suppressed at $x = 0.9$ ($x = 0.2$) \cite{Adachi-prb-2025}. 
Estimation of the spin susceptibility suggests that the unconventional pairing mechanism could work in this system \cite{Furutani-jpsj-2023}. 
We therefore may anticipate that the pairing symmetry is chiral $d$-wave at $x = 1$, and is changed to another pairing state 
without TRSB by decreasing $x$.
 
In this paper, we pursue the symmetry-changing scenario mentioned above. We utilize effective tight-binding models obtained from the first principles calculations for  $\rm{BaPtAs}_{1-\it{x}}\rm{Sb}_{\it{x}}$ with $x = 0$ and $x = 1$. We find that the chiral $d$-wave state with TRSB is most stable at $x = 1$, whereas the states without TRSB, such as nodal $f$-wave or conventional $s$-wave states, are competitive at $x = 0$.
Our results are consistent with $\mu$SR experimental results in terms of changing of pairing symmetry with respect to varying $x$ \cite{Adachi-prb-2025}.

\section{\label{sec:level2}Electronic Structures and Effective Tight-binding Models}

We perform first-principles calculations to obtain the electronic structure of pristine $\rm{BaPtAs}$ ($x = 0$) and $\rm{BaPtSb}$ ($x = 1$) using the Quantum ESPRESSO Package based on the plane wave pseudopotential method \cite{Giannozzi-jpcm-2009,Giannozzi-jpcm-2017}. 
The electronic
wave functions are expanded in plane wave functions with cutoff energy 80 Ry. The 24$\times$24$\times$24 (48$\times$48$\times$48) $\bm{k}$-point meshes for the sampling of the Brillouin Zone (BZ) in  self-consistent field  (non self-consistent field) calculations and the Perdew-Burke-Ernzerhof exchange-correlation functional are employed \cite{Perdew-prl-1996}.
The experimentally obtained lattice constants are $a$= 4.308 $\mathring{\text{A}}$ and $c$ = 4.761 $\mathring{\text{A}}$ \;for $\rm{BaPtAs}$ and $a$ = 4.535 $\mathring{\text{A}}$ and $c$ = 4.884 $\mathring{\text{A}}$ for $\rm{BaPtSb}$ \cite{Wenski-1986, Kudo-jpsj-2018-BaPtAs}.
The lattice parameters obtained by the optimization with including spin-orbit coupling (SOC) are $a$ = 4.372 $\mathring{\text{A}}$ and $c$ = 4.836 $\mathring{\text{A}}$ for $\rm{BaPtAs}$, $a$ = 4.607 $\mathring{\text{A}}$ and $c$ = 4.950 $\mathring{\text{A}}$ for $\rm{BaPtSb}$, therefore the differences between experimental and optimized values are within 1.7\%.
We define the primitive translation vectors as $\bm{a}_1 = a(\sqrt{3}/2,-1/2,0)$, $\bm{a}_2 = a(0,1,0)$ and $\bm{c} = c(0,0,1)$.

\begin{figure}[tb]
\includegraphics[width=\linewidth]{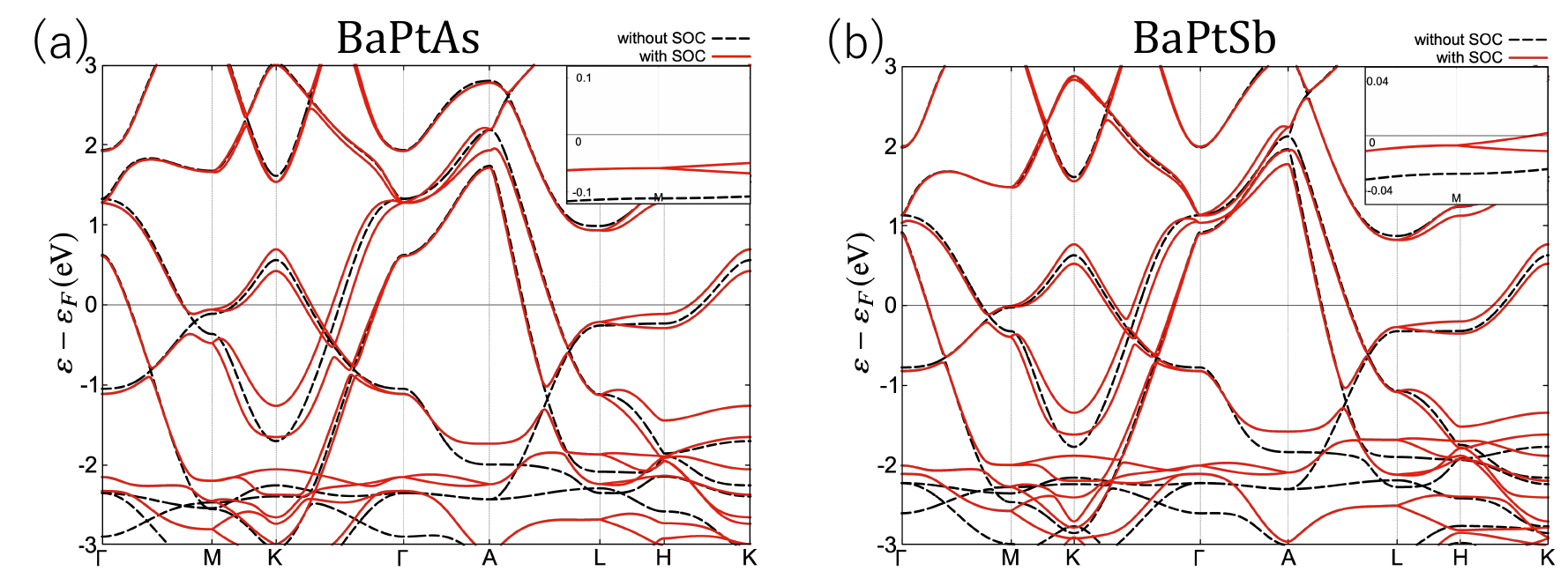}% Here is how to import EPS art
\caption{\label{fig:DFT_band}(Color online) The electronic band structures of (a) $\rm{BaPtAs}$ and (b) $\rm{BaPtSb}$. Red solid (black dashed) lines denote electronic band calculated with (without) SOC. Spin degeneracy is lifted by the ASOC. Insets show magnifications around the Fermi level near the M point (saddle point). }
\end{figure}

\begin{figure}[tb]
\includegraphics[width=\linewidth]{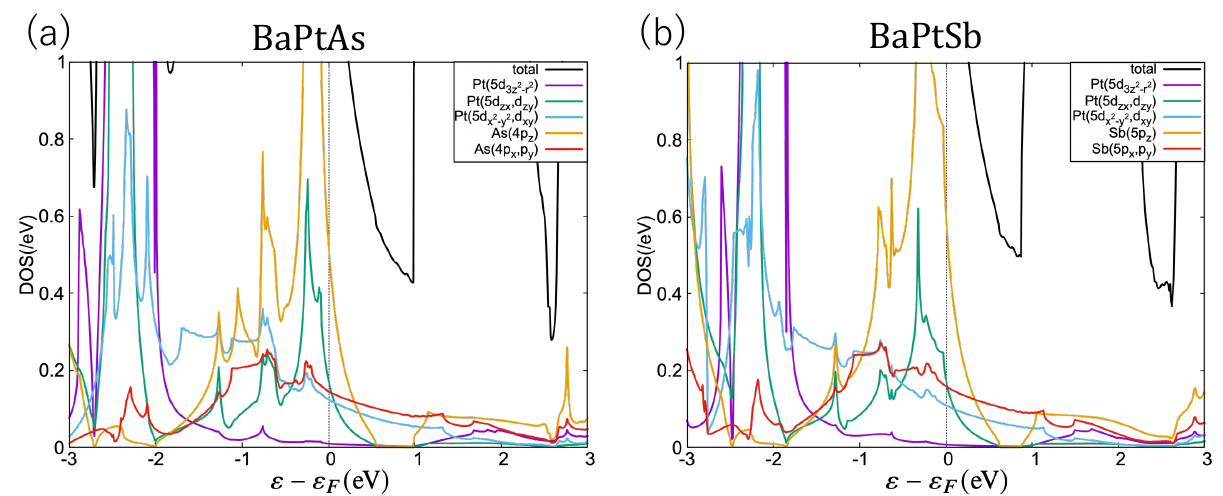}% Here is how to import EPS art
\caption{\label{fig:dos}(Color online) DOS calculated without SOC around the Fermi level of (a) $\rm{BaPtAs}$ and (b) $\rm{BaPtSb}$. }
\end{figure}

\begin{figure*}[tb]
\includegraphics[width=\linewidth]{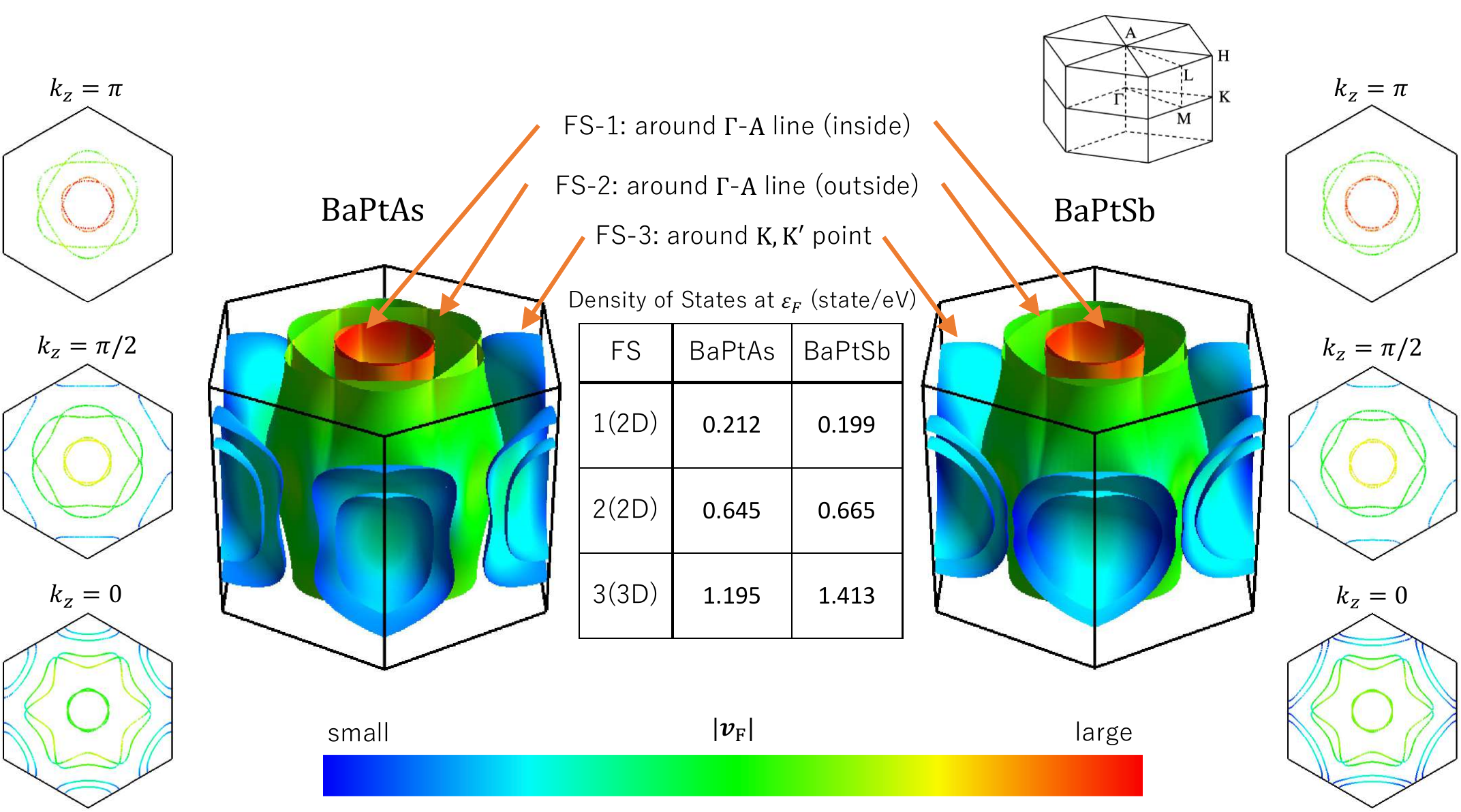}% Here is how to import EPS art
\caption{\label{fig:fs}(Color online) Fermi surfaces in 3D BZ and in 2D BZ for (left side) $\rm{BaPtAs}$ and (right side) $\rm{BaPtSb}$ visualized using Fermisurfer \cite{Kawamura-cp-2019}. The center table shows the DOS of each Fermi surfaces  that is estimated using tight-binding model mentioned in the text. The colour scale illustrates the magnitude of the Fermi velocity.}
\end{figure*}

The low-energy electronic structures are shown in Fig. \ref{fig:DFT_band}(a) for $\rm{BaPtAs}$ and (b) for $\rm{BaPtSb}$ where red solid (black dashed) lines are electronic band with (without) SOC.
Degeneracy with respect to the electron's spin is lifted due to the antisymmetric spin-orbit coupling (ASOC) attributed to the lack of inversion center in the unit cell.
An energy level at the saddle point (M point) is close to the Fermi level $\varepsilon_F$ in both compounds, yielding a van Hove singularity (VHS) that enhances the density of states (DOS) around the $\varepsilon_F$.
As highlighted in the insets, the VHS in $\rm{BaPtSb}$ is located much closer to $\varepsilon_F$ than that in $\rm{BaPtAs}$.  
These results are consistent with several previous studies \cite{Tutuncu-prb-2019, Uzunok-Intermetallics-2019, Imai-jpcs-2022, Furutani-jpsj-2023}.
Total and partial DOS calculated without SOC are shown in Fig.~\ref{fig:dos}(a) for $\rm{BaPtAs}$ and (b) for $\rm{BaPtSb}$.
Around $\varepsilon_F$, the pnictogen $4p_z$ ($5p_z$) orbital and $\rm{Pt}$ $5d_{xz}$ and $5d_{yz}$ orbitals are dominant in the DOS in $\rm{BaPtAs}$ ($\rm{BaPtSb}$).
The electronic orbitals hybridize strongly and the electronic structures around the $\varepsilon_F$ consist of two groups, group A: \{$\rm{Pt}$ : 5($d_{x^2-y^2}, d_{xy}$), $\rm{As/Sb}$ : 4($p_{x}, p_{y}$) / 5($p_{x}, p_{y}$) \}
 and group B: \{$\rm{Pt}$ : 5($d_{xz}, d_{yz}$), $\rm{As/Sb}$ : 4$p_{z}$ / $5p_{z}$ \} \cite{Imai-jpcs-2022, Furutani-jpsj-2023}.
The orbitals of group A (B) construct quasi-two (three)-dimensional Fermi Surfaces. 
As shown in Fig.~\ref{fig:fs}, we refer to the inner cylindrical Fermi surfaces around $\Gamma$ point, the outer ones, and the three-dimensional ones around the zone boundary  as FS-1, FS-2, and FS-3, respectively. 
The center table in Fig.~\ref{fig:fs} represents the DOS of each Fermi surfaces  at $\varepsilon_F$. In each compound, FS-3 has dominant contribution to the DOS. 
The FS-3 of $\rm{BaPsSb}$ lies closer to the saddle point M, whereas that of $\rm{BaPsAs}$ extends predominantly along the $k_z$-axis.

\begin{figure}[tb]
\includegraphics[width=\linewidth]{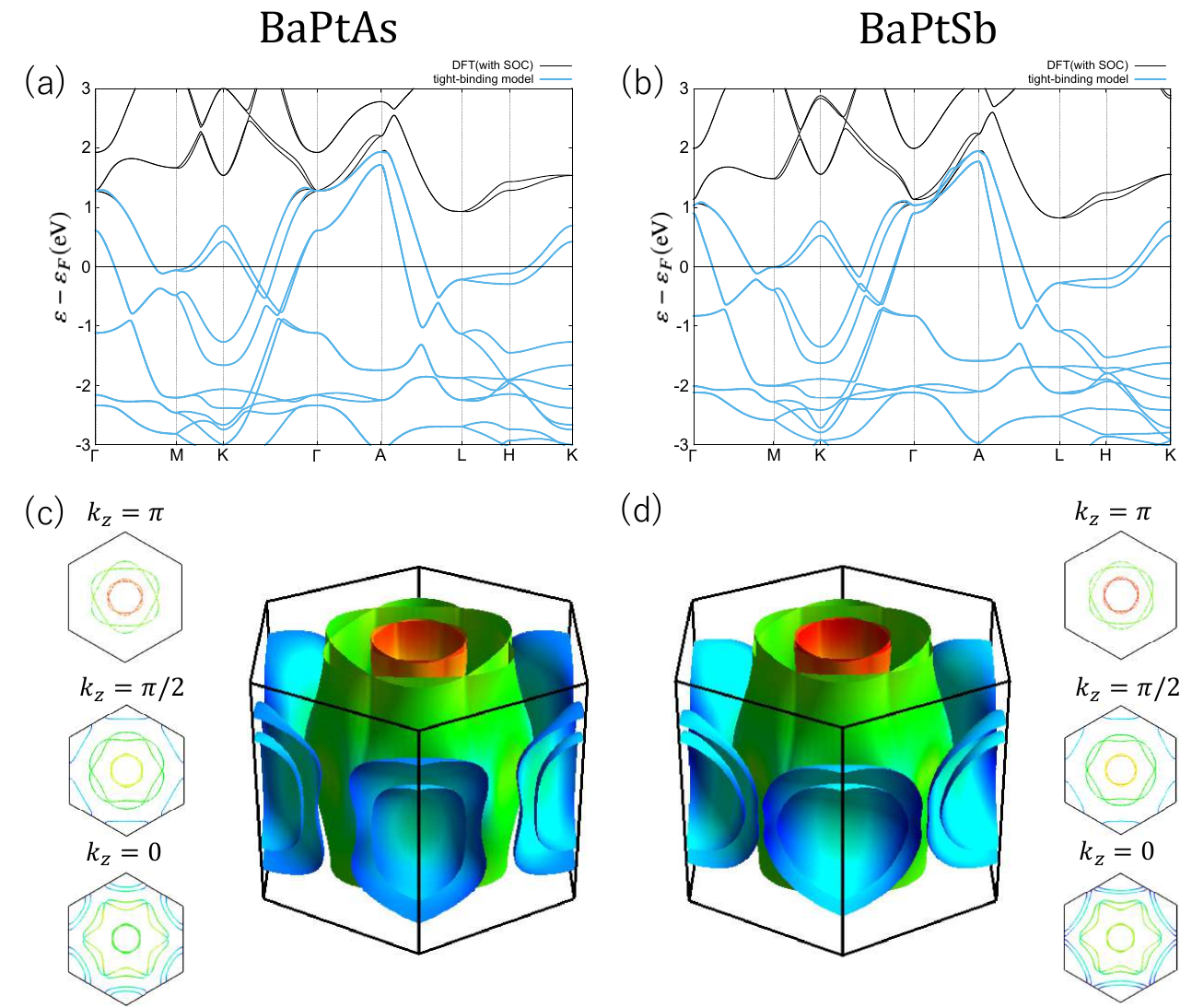}% Here is how to import EPS art
\caption{\label{fig:tb}(Color online) Energy dispersion of effective tight-binding models of (a) $\rm{BaPtAs}$ and (b) $\rm{BaPtSb}$, and corresponding Fermi surfaces  (c) $\rm{BaPtAs}$ and (d) $\rm{BaPtSb}$.}
\end{figure}

In order to analyse the superconducting state of these compounds, we construct effective tight-binding models based on maximally localized Wannier functions using Wannier90 \cite{Marzari-prb-1997,Souza-prb-2001,Pizzi-jpcm-2020}.
To reconstruct energy bands around $\varepsilon_F$, $\rm{Pt}$ $5d$ ($d_{3z^2-r^2}$, $d_{xz}, d_{yz}, d_{x^2-y^2}, d_{xy}$) and $\rm{As}$ $4p$ / $\rm{Sb}$ $5p$ ($p_z$, $p_x$, $p_y$) are chosen.
The energy dispersions of the constructed 16 orbitals effective tight-binding models (8 orbitals for each spin degrees of freedom) are shown in Fig.~\ref{fig:tb}(a) and (c). 
We find that the largest in-plane nearest-neighbor interatomic hopping is
$|t^{p_x/p_y:d_{xy}/d_{x^2-y^2}}|$, which amounts to 0.877 eV (0.769 eV) for ${\rm BaPtAs}$ (${ \rm BaPtSb}$).
The second-largest hopping is given by $|t^{p_z:d_{xz}/d_{yz}}|$, with values of 0.512 eV (0.455 eV) for ${\rm BaPtAs}$ (${ \rm BaPtSb}$), respectively.
Fermi surfaces derived from effective tight-binding models are shown in Fig.~\ref{fig:tb}(b) and (d), and the results from these models agree well with DFT calculations.
A commonly used form of the ASOC $\bm{g}$-vector in this system is
\begin{align}
    \bm{g}  \propto \hat{\bm{z}}\sigma \sum_{i=1}^3 \sin{(\bm{k}\cdot \bm{a}_i)}  \label{eq:ASOC}
\end{align}
where $\bm{a}_3\equiv-\bm{a}_1-\bm{a}_2$ \cite{Sigrist-jpsj-2014, Kudo-jpcs-2022, Furutani-jpsj-2023}.
The ASOC like Eq.(\ref{eq:ASOC}) respect $\sigma$ conservation and the Kramers pair $\ket{\bm{k},\sigma=+\hbar/2}$ and $\ket{\bm{k},\sigma=-\hbar/2}$ exists in the spin-orbit split band.
The effective Hamiltonian around the $\varepsilon_F$ is expressed as follows:
\begin{align}
     H_0&=\sum_{\bm{k},b,\sigma}\xi_{\bm{k},\sigma}^{(b)}c^{\dag}_{\bm{k},b,\sigma}c_{\bm{k},b,\sigma}\label{eq:normal-hamiltonian} 
\end{align}
where $c_{\bm{k},b,\sigma}$ is the annihilation operator of an electron in $b$-th Fermi surfaces  with wavenumber $\bm{k}$ and pseudo spin $\sigma$.
The energy dispersion of $b$-th Fermi surfaces  measured from $\varepsilon_F$  is denoted as $\xi_{\bm{k},\sigma}^{(b)}$.

\section{\label{sec:level3}Superconducting Hamiltonian and Gap Equations}

\begin{table*}[tb]
\begin{center}
\caption{Basis functions in the point group $D_{3h}$ \cite{Kudo-jpcs-2022}. 
The $\bm{k}$-space bases are $e_{\bm{k}}=\sum_{n=1}^{3} \cos{(\bm{k}\cdot\bm{a}_n)}$, $e^{+}_{\bm{k}}=\sum_{n=1}^{3} \omega^{n-1}\cos{(\bm{k}\cdot \bm{a}_n)}$, $e^{-}_{\bm{k}}=(e^{+}_{\bm{k}})^*$, $o_{\bm{k}}=\sum_{n=1}^{3} \sin{(\bm{k}\cdot \bm{a}_n)}$, $\omega=e^{i\frac{2}{3}\pi}$, $o^{+}_{\bm{k}}=\sum_{n=1}^{3}\omega^{n-1} \sin{(\bm{k}\cdot \bm{a}_n)}$ and $o^{-}_{\bm{k}}=(o^{+}_{\bm{k}})^*$ and $\omega=e^{i\frac{2}{3}\pi}$, and $\hat{\bm{\sigma}}=(\hat{\sigma}_x,\hat{\sigma}_y,\hat{\sigma}_z)$. }
\label{D3h-gap-function-table}
\begin{tabular}{c||ccc}    \hline 
Irrep($\Gamma$)  & range($r$)   & $i\Delta_s^{\Gamma,r,m}\psi_{\bm{k}}^{\Gamma,r,m}\hat{\sigma_y}$ &  $i\Delta_t^{\Gamma,r,m}\bm{d}_{\bm{k}}^{\Gamma,r,m}\cdot \hat{\bm{\sigma}}\hat{\sigma}_y$ \\ \hline \hline
$A_1'$ & on                     &   $\psi_{\bm{k}}^{A_1',\text{on},1}=1$                      &        -        \\
& ipn            &   $\psi_{\bm{k}}^{A_1',\text{ipn},1}=e_{\bm{k}}$             &        $\bm{d}_{\bm{k}}^{A_1',\text{ipn},1}=o_{\bm{k}}\hat{\bm{z}}$        \\
& opn        &   $\psi_{\bm{k}}^{A_1',\text{opn},1}=\cos{(\bm{k}\cdot \bm{c})}$              &        -        \\ \hline
$A_1''$  & ipn            &  -                                             &       $\bm{d}_{\bm{k}}^{A_1'',\text{ipn},1}=\omega{o}_{\bm{k}}^{+}(-\hat{\bm{x}}+i\hat{\bm{y}})+\omega{o}_{\bm{k}}^{-}(\hat{\bm{x}}+i\hat{\bm{y}})$         \\
& opn        &  -                                             &      $\bm{d}_{\bm{k}}^{A_1'',\text{opn},1}=\sin{(\bm{k}\cdot \bm{c})}\hat{\bm{z}} $       \\  \hline
$A_2''$    & ipn            &  -                                             &       $\bm{d}_{\bm{k}}^{A_2'',\text{ipn},1}=\omega{o}_{\bm{k}}^{+}(-\hat{\bm{x}}+i\hat{\bm{y}})+\omega^*{o}_{\bm{k}}^{-}(\hat{\bm{x}}+i\hat{\bm{y}})$         \\  \hline
$E'$    & ipn            &   $\psi_{\bm{k}}^{E',\text{ipn},1}={e}^{+}_{\bm{k}}$, \quad     $\psi_{\bm{k}}^{E',\text{ipn},2}={e}^{-}_{\bm{k}}$                                   &        $\bm{d}_{\bm{k}}^{E',\text{ipn},1}={o}^{+}_{\bm{k}}\hat{\bm{z}}$ , \quad $\bm{d}_{\bm{k}}^{E',\text{ipn},2}={o}^{-}_{\bm{k}}\hat{\bm{z}}$  \\ 
&  opn       &                                       -          &    $\bm{d}_{\bm{k}}^{E',\text{opn},1}=\sin{(\bm{k}\cdot \bm{c})}(-\hat{\bm{x}}+i\hat{\bm{y}})$ , \quad    $\bm{d}_{\bm{k}}^{E',\text{opn},2}=\sin{(\bm{k}\cdot \bm{c})}(\hat{\bm{x}}+i\hat{\bm{y}})$  \\ \hline
$E''$     & ipn            &  -                                             &       $\bm{d}_{\bm{k}}^{E'',\text{ipn},1}={o}_{\bm{k}}(-\hat{\bm{x}}+i\hat{\bm{y}})$,\quad  $\bm{d}_{\bm{k}}^{E'',\text{ipn},2}={o}_{\bm{k}}(\hat{\bm{x}}+i\hat{\bm{y}})$         \\
&                             &  -                                             &       $\bm{d}_{\bm{k}}^{E'',\text{ipn},1'}={o}^{+}_{\bm{k}}(\hat{\bm{x}}+i\hat{\bm{y}})$,\quad  $\bm{d}_{\bm{k}}^{E'',\text{ipn},2'}={o}^{-}_{\bm{k}}(\hat{\bm{x}}-i\hat{\bm{y}})$         \\ \hline
\hline
\end{tabular}
\end{center}
\end{table*}

\begin{figure}[tb]
\includegraphics[width=\linewidth]{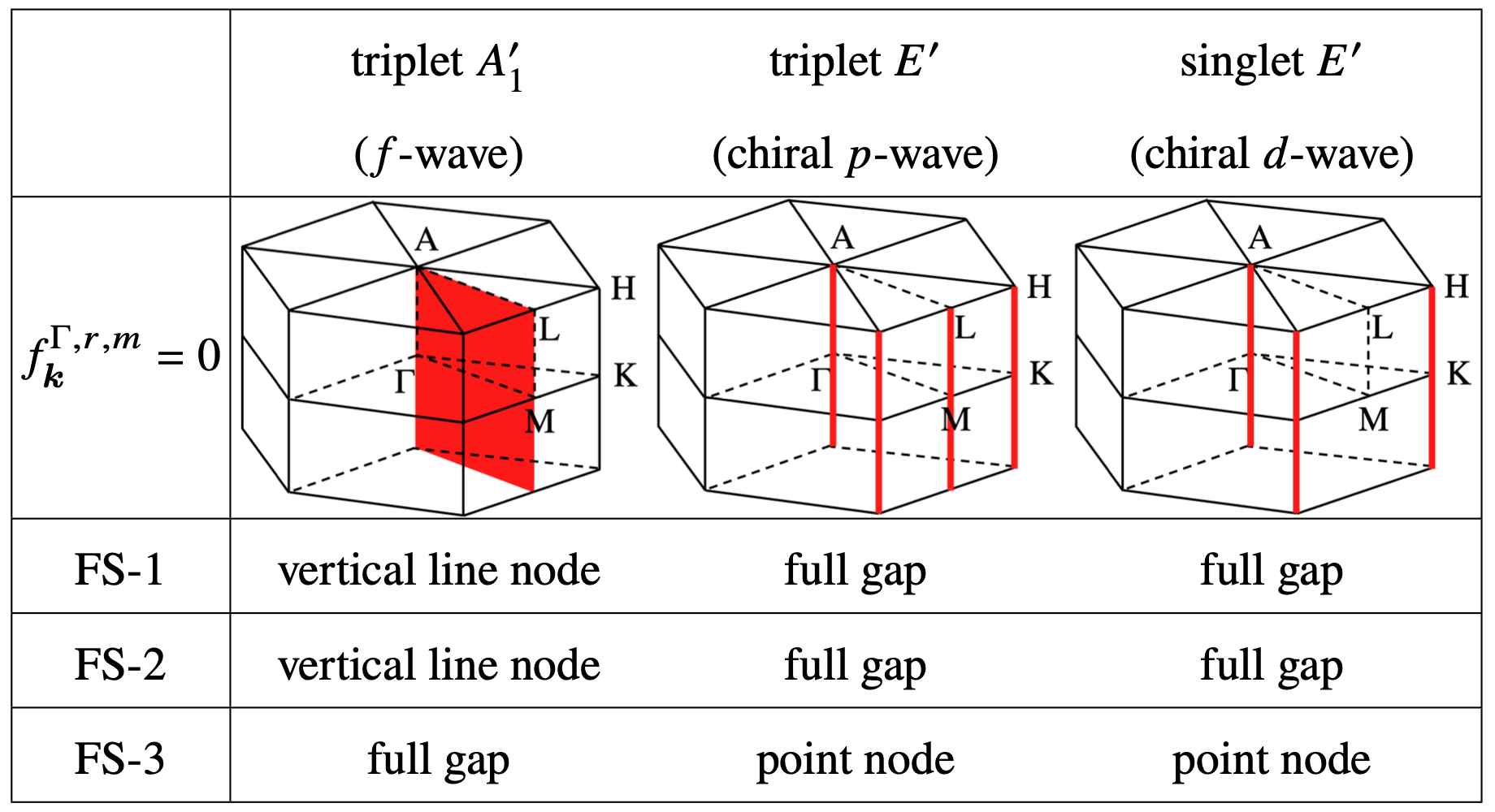}% Here is how to import EPS art
\caption{\label{fig:node-structure} (Color online) Nodal structure of the quasiparticle excitation spectrum on each Fermi surface. The first row displays the zeros of the basis functions $f_{\bm{k}}^{\Gamma,r,m}$ (red lines/planes). The excitation nodes are located at the intersect of the zero of $f_{\bm{k}}^{\Gamma,r,m}$ and the Fermi surfaces in Fig.~\ref{fig:fs}. }
\end{figure}

The pairing interaction term is
\begin{align}
    H_{\text{int}}=\frac{1}{2}&\sum_{\bm{k},\bm{k}',b,b',\sigma_1,\sigma_2,\sigma_3,\sigma_4}
    V_{\sigma_1,\sigma_2,\sigma_3,\sigma_4}^{(b,b')}(\bm{k};\bm{k}') \notag \\
    &\times
    c^{\dag}_{\bm{k},b,\sigma_1}c^{\dag}_{-\bm{k},b,\sigma_2}
    c_{-\bm{k}',b',\sigma_3}c_{\bm{k}',b',\sigma_4}, \label{eq:super-hamiltonian}
\end{align}
where $V_{\sigma_1,\sigma_2,\sigma_3,\sigma_4}^{(b,b')}(\bm{k};\bm{k}')$ denotes the pair-hopping matrix elements.
Here we suppose intra- and inter-band pairing with zero center-of-mass momentum.
The pairing basis functions $\psi_{\bm{k}}^{\Gamma,r,m}$ and $\bm{d}_{\bm{k}}^{\Gamma,r,m}$ for spin-singlet and triplet channels are listed in Table~\ref{D3h-gap-function-table}.
The label $m = 1, \cdots , N$ shows the components of the $N$-dimensional irreducible representation $\Gamma$, and $r$ denotes the interaction range in the real space.
The interaction range $r$ is restricted to on-site (on), in-plane nearest-neighbour (ipn) and out-of-plane nearest-neighbour sites (opn).
As shown in Table~\ref{D3h-gap-function-table}, the spin-singlet and spin-triplet states coexist in some irreducible representations.
Although a mixing of singlet and triplet states belonging to a same irreducible representation is possible, we neglect this effect, since it is generally expected to be small.
We adopt the normalization
\begin{align}
    \frac{1}{\Omega_{\text{BZ}}} \int_{\text{BZ}}d^3\bm{k} |\psi_{\bm{k}}^{\Gamma,r,m}|^2&=1\\
    \frac{1}{\Omega_{\text{BZ}}} \int_{\text{BZ}}d^3\bm{k} |\bm{d}_{\bm{k}}^{\Gamma,r,m}|^2&=1,
\end{align}
where $\Omega_{\text{BZ}}$ is BZ volume.
As noted in Sec. \ref{sec:level2}, the systems approximately conserve $S_z$.
Therefore, we restrict our analysis to pairing states with $S_z=0$ (Table \ref{D3h-gap-function-table-sz0}).
Using these bases, the effective interaction is
\begin{align}
    &V_{\sigma_1,\sigma_2,\sigma_3,\sigma_4}^{(b,b')}(\bm{k};\bm{k}')\notag \\
    &\quad=-\sum_{\Gamma,r,m}
\varg_{\text{s},\Gamma,r,m}^{(b,b')}\psi_{\bm{k}}^{\Gamma,r,m}\psi_{\bm{k}'}^{\Gamma,r,m*}(i\hat{\sigma}_y)_{\sigma_1\sigma_2}(i\hat{\sigma}_y)^{\dag}_{\sigma_3\sigma_4}\notag\\
&\qquad-\sum_{\Gamma,r,m}
\varg_{\text{t},\Gamma,r,m}^{(b,b')}d_{z,\bm{k}}^{\Gamma,r,m}d_{z,\bm{k}'}^{\Gamma,r,m*}(i\hat{\sigma}_z\hat{\sigma}_y)_{\sigma_1\sigma_2}(i\hat{\sigma}_z\hat{\sigma}_y)^{\dag}_{\sigma_3\sigma_4}
\end{align}
where $\varg_{\text{s},\Gamma,r,m}^{(b,b')}$ ($\varg_{\text{t},\Gamma,r,m}^{(b,b')}$) is the 
pair hopping of the $m$-th component of an irreducible representation $\Gamma$ between $b$- and $b'$-th band. 
At $k_{\text{B}}T=0$, the gap equation is expressed as follows:
\begin{align}
    \Delta^{(b)}=\sum_{b',\sigma}\varg_{s/t,\Gamma,r,m}^{(b,b')} \frac{\Delta^{(b')}}{\Omega_{BZ}}\int_{\Lambda}
    d^3\bm{k} \frac{|f_{\bm{k}}^{\Gamma,r,m}|^2}{2\sqrt{\xi_{\sigma,\bm{k}}^{(b')2}+\Delta^{(b')2}|f_{\bm{k}}^{\Gamma,r,m}|^2}}, \label{eq:gap-equation}
\end{align}
where $\int_{\Lambda} d^3\bm{k}$ denotes the momentum integral over the states
over the cut-off energy range, and  $f_{\bm{k}}^{\Gamma,r,m}$ represents the orbital part of basis function in the Table~\ref{D3h-gap-function-table-sz0}.
Fig.~\ref{fig:node-structure} summarizes the nodal structure of the states appeared in Fig.~\ref{fig:phase-diagram}.
The $f$-wave state has symmetry-protected vertical line nodes on FS-1, -2.
The chiral $p_x+ip_y$- and chiral $d_{x^2-y^2}+id_{xy}$-wave states are fully gapped on cylindrical FS-1, -2 and has point nodes on three-dimensional FS-3.

\begin{table}[tb]
\begin{center}
 \caption{Basis functions of $S_z = 0$ ($z$-component of total spin is zero) pairing of Table~\ref{D3h-gap-function-table}. }
\label{D3h-gap-function-table-sz0}
\begin{tabular}{c||ccc}    \hline 
 Irrep($\Gamma$)  & Range   & $i\Delta_s^{\Gamma,r,m}\psi_{\bm{k}}^{\Gamma,r,m}\hat{\sigma_y}$ &  $i\Delta_t^{\Gamma,r,m}\bm{d}_{\bm{k}}^{\Gamma,r,m}\cdot \hat{\bm{\sigma}}\hat{\sigma}_y$ \\ \hline \hline
$A_1'$ & on                     &   $\psi_{\bm{k}}^{A_1',\text{on},1}=1$                      &        -        \\
& ipn            &   $\psi_{\bm{k}}^{A_1',\text{ipn},1}=e_{\bm{k}}$             &        $\bm{d}_{\bm{k}}^{A_1',\text{ipn},1}=o_{\bm{k}}\hat{\bm{z}}$        \\
& opn        &   $\psi_{\bm{k}}^{A_1',\text{opn},1}=\cos{(\bm{k}\cdot \bm{c})}$              &        -        \\\hline
 $A_1''$ & opn        &  -                                             &      $\bm{d}_{\bm{k}}^{A_1',\text{opn},1}=\sin{(\bm{k}\cdot \bm{c})}\hat{\bm{z}} $       \\   \hline
$E'$    & ipn            &\begin{tabular}{c} $\psi_{\bm{k}}^{E',\text{ipn},1}={e}^{+}_{\bm{k}}$, \\ \quad     $\psi_{\bm{k}}^{E',\text{ipn},2}={e}^{-}_{\bm{k}}$ \end{tabular}&\begin{tabular}{c} $\bm{d}_{\bm{k}}^{E',\text{ipn},1}={o}^{+}_{\bm{k}}\hat{\bm{z}}$, \\  \quad $\bm{d}_{\bm{k}}^{E',\text{ipn},2}={o}^{-}_{\bm{k}}\hat{\bm{z}}$ \end{tabular}
                                      \\ \hline
\hline
\end{tabular}
\end{center}
\end{table}

\section{\label{sec:level4}Phase Diagrams}

\begin{figure}[tb]
\includegraphics[width=\linewidth]{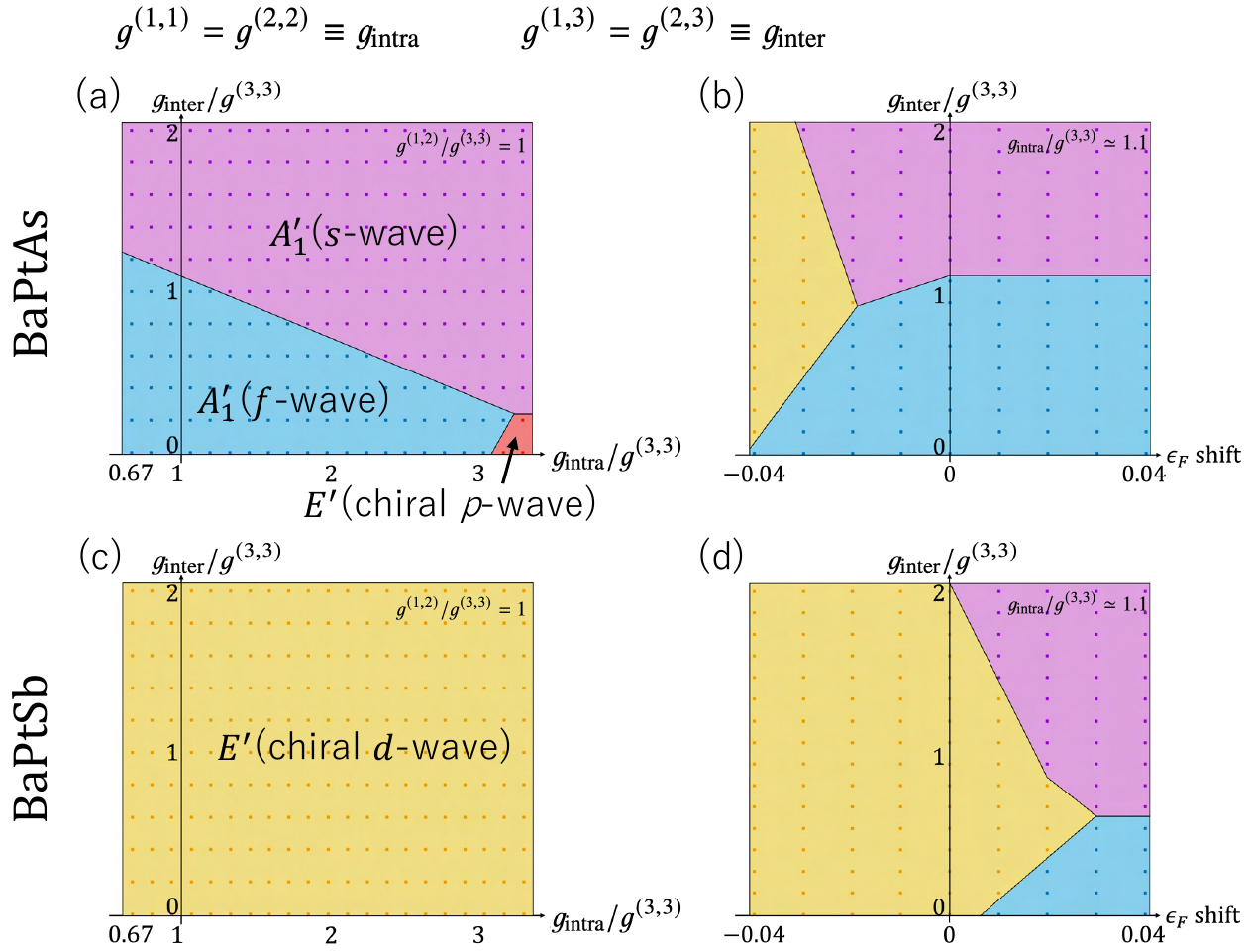}% Here is how to import EPS art
\caption{\label{fig:phase-diagram} (Color online)
Phase diagrams obtained by solving the gap equation (\ref{eq:gap-equation}) and comparing the condensation energy (\ref{eq:con}) for each irreducible representation; at each point the illustrated state is the one with the largest $|E_{\text{con}}|$. 
We neglect the irreducible representation dependence of the coupling constants, i.e., $\varg_{s/t,\Gamma,r,m}^{(b,b')}=\varg^{(b,b')}$.
The coupling-constant space is 6-dimensional, we reduce dimensionality by imposing
$\varg^{(1,1)}=\varg^{(2,2)} \equiv \varg_{\text{intra}}$ and $\varg^{(1,3)}=\varg^{(2,3)}\equiv \varg_{\text{inter}}$.
Panels (a) and (c): the in-plane axes are $\varg_{\text{intra}}/\varg^{(3,3)}$ and $\varg_{\text{inter}}/\varg^{(3,3)}$, and $\varg^{(1,2)}/\varg^{(3,3)}=1$ is chosen (we confirmed a only slight change with varying the value $\varg^{(1,2)}/\varg^{(3,3)}$).
Panels (b) and (d): the in-plane axes are $\varepsilon_F$ and $\varg_{\text{inter}}(=\varg^{(1,2)})/\varg^{(3,3)}$, and $\varg_{\text{intra}}/\varg^{(3,3)}\simeq 1.1$ is chosen.}
\end{figure}

We solve the zero-temperature gap equation Eq. (\ref{eq:gap-equation}) and evaluate the condensation energy with a cut-off $0.01$ eV (The bandwidth is 3--4 eV. And we confirm the following results are robust about cut-off size.) :
\begin{align}
    E_{\text{con}}
    &=\braket{H}-{H}_{\Delta=0} \notag \\
    &=\sum_{b,\sigma} \frac{1}{2\Omega_{BZ}}\int_{\Lambda}
    d^3\bm{k} \Bigg[ -\sqrt{\xi_{\sigma,\bm{k}}^{(b)2}+\Delta^{(b)2}|f_{\bm{k}}^{\Gamma,r,m}|^2}\notag \\
&\qquad+|\xi_{\sigma,\bm{k}}^{(b)}|+\frac{\Delta^{(b)2}|f_{\bm{k}}^{\Gamma,r,m}|^2}{2\sqrt{\xi_{\sigma,\bm{k}}^{(b)2}+\Delta^{(b)2}|f_{\bm{k}}^{\Gamma,r,m}|^2}}\Bigg]. \label{eq:con}
\end{align}
We construct phase diagrams in the parameter space of coupling constants.
To reduce the parameter space, we set $\varg_{s/t,\Gamma,r,m}^{(b,b')}=\varg^{(b,b')}$, and $\varg^{(1,1)}=\varg^{(2,2)} \equiv \varg_{\text{intra}}$ and $\varg^{(1,3)}=\varg^{(2,3)}\equiv \varg_{\text{inter}}$.
See, also the caption in Fig.~\ref{fig:phase-diagram}.
The illustrated pairing symmetry in Fig.~\ref{fig:phase-diagram} is the most stable pairing symmetries  identified as the state with the largest condensation energy in all the states listed in Table~\ref{D3h-gap-function-table-sz0}.

First, we examine the results in Fig.~\ref{fig:phase-diagram} (a) and (c).
As shown in Fig.~\ref{fig:node-structure}, the excitation spectra of the $f$-wave state is fully gapped on three-dimensional FS-3 and have line nodes on cylindrical FS-1, -2. 
On the other hand, the chiral $p$-, $d$-wave states are fully gapped on FS-1, -2 and possess polar point nodes on FS-3. 
While the amplitudes of the $f$- and chiral $d$-wave states increase near FS-3, a significant difference exists: near the M point (a saddle point), the amplitude of chiral $d$-wave ($f$-wave) state becomes maximum (is absent).
Therefore, in $\rm{BaPtSb}$, where FS-3 extends to the vicinity of the M point, the chiral $d$-wave state becomes stable, while in $\rm{BaPtAs}$ with a relatively shrunk FS-3, the $f$-wave state becomes stable.
The chiral $p$-wave state stabilizes when the coupling constant on cylindrical Fermi surfaces  $\varg_{\text{intra}}$ dominates, as the amplitude of the gap function reaches its maximum near the cylindrical Fermi surface FS-1, -2. 
The conventional $s$-wave state, which is fully gapped at all Fermi surfaces, competes with these unconventional states.
Since the $f$-wave state possesses line nodes on FS-1, -2, large interband coupling constants adversely affect its stabilization. Therefore, in $\rm{BaPtAs}$, $s$-wave state exhibit superior stability when the interband coupling constants are large.

We then discuss the results of Fig.~\ref{fig:phase-diagram} (b) and (d).
The difference between these two materials can be naively understood as a difference of the Fermi level.
Indeed, the Fermi surface of $\rm{BaPtSb}$ closely resembles that of $\rm{BaPtAs}$ under a Fermi level shift with hole doping. In $\rm{BaPtAs}$ compared with $\rm{BaPtSb}$, the three-dimensional FS-3 is smaller and located farther from the M--L line, and the associated VHS lies farther from the Fermi level. 
In contrast, the cylindrical FS-1, -2 are similar in the two compounds.
Fig.~\ref{fig:phase-diagram}(b) and (d) show phase diagrams for a rigidly shifted Fermi level in $\rm{BaPtAs}$ and $\rm{BaPtSb}$, respectively. 
The two phase diagrams are highly similar, except for their dependence on the interband coupling constant.
Their close resemblance indicates that shifting Fermi surface toward the M--L line -- thereby bringing the Fermi level closer to the VHS  -- favors the chiral $d_{x^2-y^2}+id_{xy}$-wave state. 
Because the amplitude of the basis function of the chiral $d$-wave state is large in the M--L neighborhood, the contribution from FS-3 near this region gains weight in the gap equation.
That increases the condensation energy and expands the chiral $d$-wave domain in the phase-diagrams.
It is expected that the hole doping stabilizes the chiral $d$-wave state by lowering the Fermi level of $\rm{BaPtAs}$.

In the previous study by Furutani et al.~\cite{Furutani-jpscf-2023}, the contribution from the Pt ($d_{xz}$, $d_{yz}$) orbitals was not taken into account.
In the present work, we explicitly include these orbitals, since they provide a non-negligible contribution to the three-dimensional Fermi surface (FS-3).
However, as shown in Sect.~\ref{sec:level2}, the contribution of the Pt ($d_{xz}$, $d_{yz}$) orbitals in ${\rm BaPtSb}$ is smaller than that in ${\rm BaPtAs}$.
We therefore conclude that these orbitals do not play a significant role in stabilizing the chiral $d$-wave state in ${\rm BaPtSb}$.
Instead, the stabilization of the chiral $d$-wave state is essentially governed by the fact that the FS-3 in ${\rm BaPtSb}$ lies closer to the band saddle point at the $M$ point than that in ${\rm BaPtAs}$, as discussed above.

\section{\label{sec:level5}Conclusion}

We have investigated the stable pairing symmetry of  $\rm{BaPtAs}_{1-\it{x}}\rm{Sb}_{\it{x}}$  with an ordered honeycomb network by combining tight-binding models derived by first-principles calculations. By solving the zero-temperature gap equation and evaluating the condensation energy in the coupling-constant space, we constructed phase diagrams. The results show that in BaPtSb ($x = 1$) the chiral $d_{x^2-y^2}+id_{xy}$ state is most stable over a broad parameter range, whereas in BaPtAs ($x = 0$) states without TRSB are competitive.
Our present results are consistent with $\mu$SR experimental results \cite{Adachi-prb-2025}.
To settle the pairing symmetry across $\rm{Sb}$-concentration $x$, further measurements -- Knight shift, spin-lattice relaxation rate $T_1^{-1}$, temperature-dependent specific heat, in-plane anisotropy of the upper critical field, together with polar Kerr effect -- would be highly informative.
Furthermore, it  is predicted that hydrostatic pressure can alter the Fermi surface, bringing the VHS closer to the Fermi level, which may contribute to the stabilization of the chiral $d$-waves \cite{Furutani-jpscf-2023}.
The present results demonstrate that this material has intriguing properties, encouraging further detailed studies in the future.

\begin{acknowledgments}
T.I. is grateful to H. Ueki, R. Oiwa and T. Miki for useful discussions. 
This work was partially supported by JST SPRING Grant Number JPMJSP2152, and by JSPS KAKENHI (Grant Number JP22H01182, JP23K22453, and JP24K21531).
\end{acknowledgments}

\bibliographystyle{apsrev4-1}
\bibliography{71800}% Produces the bibliography via BibTeX.

\end{document}